# Multi-channel Speech Enhancement with 2-D Convolutional Time-frequency Domain Features and a Pre-trained Acoustic Model

Quandong Wang, Junnan Wu, Zhao Yan, Sichong Qian, Liyong Guo, Lichun Fan, Weiji Zhuang, Peng Gao, Yujun Wang

Xiaomi Corporation, Beijing, China

E-mail: {wangquandong,wujunnan1,yanzhao,qiansichong,guoliyong,fanlichun1,zhuangweiji, gaopeng11,wangyujun}@xiaomi.com

*Abstract*— We propose a multi-channel speech enhancement approach with a novel two-stage feature fusion method and a pre-trained acoustic model in a multi-task learning paradigm. In the first fusion stage, the time-domain and frequency-domain features are extracted separately. In the time domain, the multi-channel convolution sum (MCS) and the inter-channel convolution differences (ICDs) features are computed and then integrated with the first 2-D convolutional layer, while in the frequency domain, the log-power spectra (LPS) features from both original channels and super-directive beamforming outputs are combined with a second 2-D convolutional layer. To fully integrate the rich information of multi-channel speech, i.e. time-frequency domain features and the array geometry, we apply a third 2-D convolutional layer in the second fusion stage to obtain the final convolutional features. Furthermore, we propose to use a fixed clean acoustic model trained with the end-to-end lattice-free maximum mutual information criterion to enforce the enhanced output to have the same distribution as the clean waveform to alleviate the over-estimation problem of the enhancement task and constrain distortion. On the Task1 development dataset of ConferencingSpeech 2021 challenge, a PESQ improvement of 0.24 and 0.19 is attained compared to the official baseline and a recently proposed multi-channel separation method.

## I. Introduction

With the rapid development of deep learning techniques, speech enhancement, as well as speech separation, has made a great progress in recent years. Many single channel speech separation and enhancement methods based on deep learning have been proposed, such as deep clustering [1], permutation invariant training(PIT) [2][3], time-domain audio separation network (TasNet) [4]-[6], Wavesplit [7], U-Net [8], SN-Net [9], DCCRN [10] and so on. All these methods could utilize the information well in time domain or frequency domain.

In practice, microphone arrays are commonly assigned to record multi-source multi-channel data. Once multiple microphones are available, the spatial information associated with sources can be exploited for speech enhancement. Inter-channel phase differences (IPDs) are the most commonly used spatial features and have been proven to be beneficial, especially when combining with monaural spectral features as the input feature for time-frequency (T-F) masking based methods[11]-[14]. Unfortunately, it is not straightforward to incorporate IPDs with time-domain methods as IPDs are typically extracted from frequency domain with fixed complex filters (i.e., short time Fourier transform, STFT) whose window type/length and hop size are different from the encoders used in time domain. In order to enhance the source signal from a desired direction, directional features associated with a certain direction which indicate the desired source's dominance in each T-F bin have been presented in [15][16]. However, the knowledge of the target direction is unknown in real applications, and it is difficult to accurately estimate. Spatial information can be easily utilized by frequency-domain beamforming methods naturally [17][18], which have obtained great progress when combined with deep neural network (DNN). Such DNNs are usually incorporated into the beamforming framework to estimate parameters or masks [19][20][21]. In the time domain, spatial information such as multi-channel convolution sum (MCS) and inter-channel convolution differences (ICDs), have also been demonstrated to be beneficial [22]. Although using time-domain features are effective, the enhanced performance often degrades sharply when processing unseen data, while the frequency-domain features, e.g. log-power spectra (LPS), have strong generalization ability [23]. Reference [24] has conducted deep fusion on different kinds of frequency domain features. However, the research on the fusion of time-frequency domain features is not extensive in the multi-channel enhancement literatures.

Apart from the feature extraction, the objective function has fundamental impact on the enhancement performance. While most architectures combining enhancement objective and recognition objective mainly focus on the improvement of recognition accuracy [25][26], there arise recently methods aiming to use classification objective to aid the regression objective. Reference [27] proposed to use end-to-end automatic speech recognition (ASR) training objectives to train an enhancement system without using parallel data and shows good effectiveness. In [28], the authors reformulated speech enhancement as a classification model in an ASR manner instead of a regression model. Specifically, a quantized speech prediction model was proposed to recover more realistic speech spectra against time-frequency masking approaches.

In this work, we propose a multi-task method to enhance the source signal, in which we consider the speech enhancement as the main task and the ASR as the auxiliary task. For the main task, we train a multi-channel speech enhancement model with a two-stage feature fusion method, in which we integrate time-domain features, frequency-domain features and spatial features together by three 2-D convolutional layers. For the auxiliary task, we propose to constrain the output speech distortion by applying an acoustic model trained with clean speech data using the end-to-end lattice-free maximum mutual information (LF-MMI) criterion [29][30] which is commonly used in ASR task.

The rest of this paper is organized as follows. Section 2 describes our proposed system in detail. Data generation, experimental results and analysis are presented in Section 3. Section 4 concludes the paper.

## II. PROPOSED METHOD

Fig. 1 shows the overall structure of our proposed system. Two parts are included: speech enhancement branch and acoustic model branch. We introduce them in detail as follows.

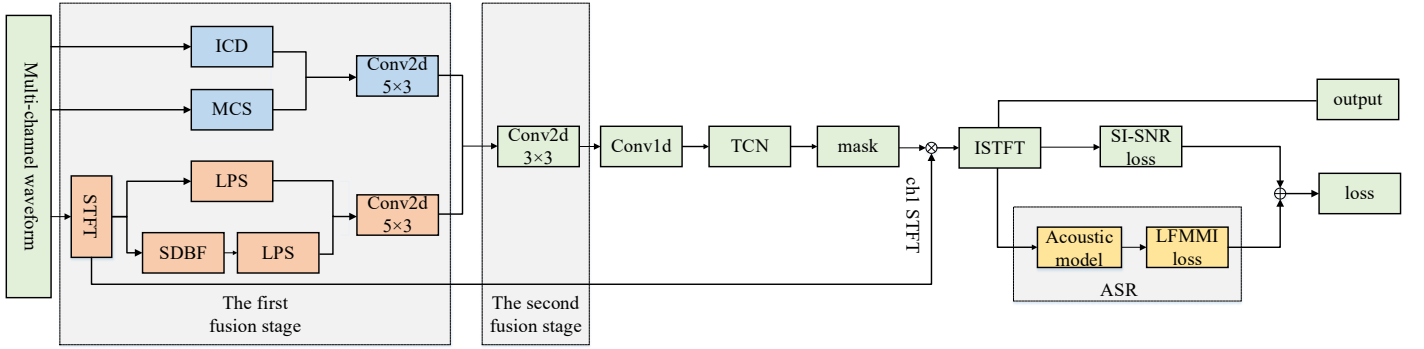

Fig. 1 The overall structure of our proposed system.

### A. Speech enhancement branch

This part is implemented to produce enhancement output waveform. The input features to the backbone network are generated by an innovative two-stage feature fusion method.

First, we extract the time-domain and frequency-domain features in the first stage of fusion. In the time domain, MCS and ICDs features are computed [22]. MCS and ICDs both take fully advantage of the inter-channel information in the time domain. MCS is designed similar to the filter-and-sum beamforming, and the $n$th ( $n=1,...,N$ ) MCS output is computed by summing up the convolution products between the $c$th channel mixture signal $x_c$ and filter $k_{cn}$ along signal channel $c$, i.e.

$$\text{MCS}_n = \sum_{c=1}^{C} x_c \otimes k_{cn}, \quad (1)$$

where $\otimes$ denotes the convolution operation. MCS utilizes the information of all input channels and is implemented by a 2-D convolutional layer, with the kernel size $(C \times L)$ for the filter $k_{cn}$, in which $C$ is the channel number and $L$ is the window length. The stride along width axis is fixed as $L/2$ in our experiments. ICDs are designed similar to IPDs from the frequency domain. The $n$th ( $n=1,...,N$ ) ICD between the $m$th channel pair is computed by the convolution products between the $m$th pair of signals and the corresponding filter $k_n^{'}$,

$$\text{ICD}_n^{(m)} = \omega_1 \cdot (x_{m_1} \otimes k_n^{'}) - \omega_2 \cdot (x_{m_2} \otimes k_n^{'}) \quad (2)$$

where $x_{m_1}$ and $x_{m_2}$ represent the signals of the $m$th pair, $\omega_1 \in \mathbb{R}^{1 \times L}$ is fixed as full ones and $\omega_2 \in \mathbb{R}^{1 \times L}$ as a learnable parameter initialized with ones. The ICD layer is implemented by a 2-D dilated convolutional layer with the kernel size $(2 \times L)$. The stride is also set to $L/2$, and the dilation is 4. Then the MCS and ICDs are integrated with the first 2-D fusion convolutional layer with kernel size $(5 \times 3)$. Note that all 2-D convolutional layers mentioned below are composed of a 2-D convolutional operation, a ReLU (rectified linear unit) activation, and a 2-D batch-norm activation, unless otherwise stated.

As stated in section I, it is necessary to explore to extract frequency-domain features in addition to time-domain features. STFT is performed to transform the signals from time domain to frequency domain, with the window size being $L$ and the hop size being $L/2$, which are the same to that in time-domain features. Then, both intra-channel and inter-channel features are employed in the frequency domain. The intra-channel features are the LPS features from original channels, while the inter-channel feature is the LPS feature of super-directive beamformer (SDBF) [31] outputs which utilize the array geometry information directly and assume spherically isotropic noise field. The SDBF is performed in uniformly distributed directions, and the output $Y$ is computed by the product of the SDBF weight vector $w_{SD}$ and multi-channel input $X$,

$$Y = w_{SD}{}^H X, \quad (3)$$

where H is the Hermitian (conjugate transpose) operator. The details of $w_{SD}$ computation is referred to [21][31]. Then a second 2-D fusion convolutional layer with kernel size $(5\times 3)$ is used to combine the above extracted intra- and inter-channel frequency features.

In the second fusion stage, a third 2-D convolutional fusion layer with kernel size $(3\times 3)$ is applied to form the final convolutional features as 2-D convolution operation can extract the local features from time and frequency domain simultaneously. Then the output is reshaped and passed to the backbone network, a dilated convolutional neural network (CNN) same to the separation module in Conv-TasNet [5], which is denoted as TCN (temporal convolutional network). A complex mask is obtained as output of TCN. Then the estimated speech spectrum is formed by multiplying the input complex spectrum of the first channel by the mask. Finally inverse STFT is used to convert the estimated spectrum into time domain and overlap-and-add method is used.

### B. Acoustic model branch

As we empirically found, neural network based speech enhancement methods suffer severely from the over-estimation problem and cause obvious distortion especially when non-stationary noise exists. In this context, we add a subsequent task of acoustic modeling to constrain the distortion of the enhanced waveform. The intuition behind is that a fixed acoustic model trained with clean data could extract the structural and detailed feature of clean speech, thus the enhancement branch output would gradually approximate the clean speech by performing the backpropagation. In the acoustic model branch, the LPS feature of the enhanced waveform is extracted and feed into the acoustic model branch whose architecture is the time delay neural network (TDNN) [32][33] designed to model both left and right context during training. Note, in the inference stage, we remove this branch to spare unnecessary computation and keep the system zero look-ahead.

### C. Loss functions

As stated above, the proposed system has two branches, thus the loss function $l_{total}$ is composed of two objectives. SI-SNR (scale-invariant signal-to-noise ratio) loss [34] $l_{enh}$ is used in the speech enhancement branch, which is one of the most commonly used loss functions in speech enhancement tasks. In the second branch, the end-to-end LF-MMI loss $l_{am}$ [29][30] is used to ensure the recognition performance because of its discriminative nature, especially when the training set is relative small. The total loss function $l_{total}$ is calculated as follows,

$$l_{total} = \alpha l_{enh} + \beta l_{am}, \quad (4)$$

$$l_{enh} = -10\log_{10}\frac{\|x_{target}\|_2^2}{\|e_{noise}\|_2^2}, \quad (5)$$

where $\alpha$ and $\beta$ are the weights for SI-SNR loss and LF-MMI loss, respectively. And $x_{target} = \langle \hat{x}, x\rangle x / \|x\|_2^2$, $e_{noise} = \hat{x} - x_{target}$, $\hat{x}$ and $x$ are the estimated source and clean waveform, respectively.

## III. EXPERIMENTAL SETUP & RESULTS ANALYSIS

### A. Data augmentation and generation

Here the data from the ConferencingSpeech 2021 challenge [35] is used to evaluate our proposed system. We focus on Task1 of the challenge: multi-channel speech enhancement with a single microphone array. In this task, speech from a single linear microphone array with non-uniform distributed microphones is acquired to perform enhancement and no future information (zero look-ahead) could be used for practical application requirement. Fig. 2 shows the experimental environment and array configuration. The array is a linear array with eight non-uniformly distributed microphones. The intervals among microphones could be referred in Fig. 2.

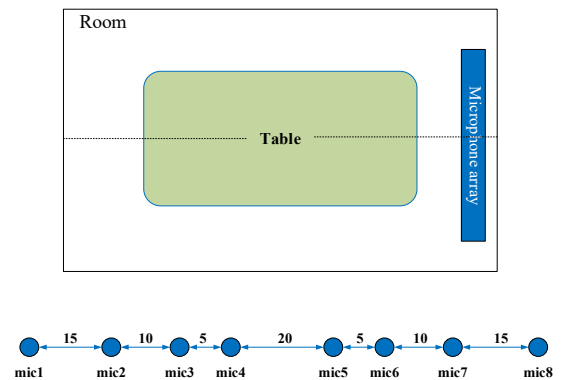

Fig. 2  The setup of microphone array in the room. The unit is centimeter.

The total duration of our training set is 122 hours for extensive experiments. We simulate 8-channel received speech by convolving single-channel clean speech and noise with their corresponding simulated multi-channel room impulse responses (RIRs) and mixing them together. The clean speech is select randomly from the public data sets (AISHELL-1 [36], AISHELL-3 [37], VCTK [38], Librispeech [39]), and the noise sources from MUSAN [40] and Audioset [41]. In order to improve the generalization of our enhancement model, we do three kinds of augmentation on the clean speech, including speed, tempo and volume perturbations, right before convolution with RIRs. The percentage of the three kinds of augmentations is set to be 1:2:3. We control the ratio of small, medium, and large rooms to be 2:3:3 to improve the dereverberation performance when processing the speech in medium and large rooms. To be specific, a room with width and length in [3m, 5m) is defined to be small, a room with width and length in [5m, 7m) is defined as medium, and that with [7m, 8m] width and length is defined as large. All the waveforms are resampled to 16 kHz.

The 8-channel development test set is generated similarly to the training set, except the data augmentation. Every utterance in the development test set is truncated to 6s. The clean speech and RIRs used in the development test set is different from that in the training set.

*B. Experimental setting*

We implemented our proposed method in Pytorch. In order to decrease the communication burden between parallel training machines, we adopted the Blockwise Model-Update Filtering (BMUF) algorithm [42] to accelerate the training progress for the proposed multi-task at the price of very small performance degradation.

For all models, the initial learning rate is set to 1e-3. And the learning rate is determined by Adam optimizer with weight-decay of 1e-5. The Hanning window size is 20ms, the hop size is 10ms, and FFT length is 512. The parameters in TCN is the same as the best setup of Conv-TasNet in [5]. When applied, four pairs of IPDs are computed from channel 1 (ch1) and ch5, ch2 and ch6, ch3 and ch7, ch4 and ch8, respectively. And the ICD features are also extracted among the four pairs of microphones. For MSC, ICD and IPD of each pair, the output dimension is $N=257$. For the SDBF module, only the first two channels are used to do beamforming on seven directions ($i\pi/8$, $i=1,2,...7$) for saving computation because more channels would lead to narrower beam width and the directions have to be increased much to cover the whole spatial space. For the proposed two-stage fusion, the first fusion Conv2D layer has the number of input and output channels $(5,8)$ with stride $(2,1)$, the second fusion Conv2D layer has $(15,8)$ with stride $(2,1)$ and the third fusion Conv2D layer has $(16,8)$ with stride $(1,1)$. When Conv1D fusion is applied, the number of output channels are 128, while the input channels are dependent on the configurations. For all convolutional operations, causal convolution is applied.

For acoustic model training, we use JIEBA [44] for word segmentation. Then the CMU dictionary and MDBG Chinese dictionary [45] are used because the training set contains both English words and Chinese characters. Nine TDNNs are adopted to construct the acoustic model. In each TDNN, there are two feed-forward layers imitating the TDNN-F structure [33], where the first one has input-dim of 1536 and output-dim of 512 while the second one has input-dim of 512 and output-dim of 1536. One final linear layer projects the TDNN output to 3920 which is the number of biphone modeling units. Practically, we first train the clean acoustic model (AM) and then fix the parameters of it when doing multi-task learning with the total loss. The loss weights for the two branches are equal, i.e., $\alpha=\beta=1.0$.

To evaluate the quality of the enhanced speech, we use four kinds of objective measures, perceptual evaluation of speech quality (PESQ) [46], short term objective intelligibility (STOI) [47][48], extended STOI (E-STOI) and SI-SNR.

*C. Architectures comparisons*

To demonstrate the effectiveness of our proposed architecture, we conducted experiments in comparisons with two baseline systems. Our first baseline system is the official baseline of the challenge. The input feature is the concatenate of the frequency-domain complex spectrum of the first channel and the four pairs of IPDs between different channels. The model structure is a 3-layer LSTM followed by a linear layer outputting a complex mask to be multiplied with the input complex spectrum. To keep consistency, we use frequency-domain features in all following experiments. Our second baseline system is developed with the same structure (TCN) to [22] except that we used single-channel LPS instead of the time-domain single-channel encoder. It concatenates four kinds of frequency and time domain features, namely, first-channel LPS, four pairs of IPDs, MCS and ICD features. It feeds the concatenated input to a Conv1D layer and then the backbone TCN module. As TCN architecture attains excellent performance as well as low computation, we use it as our backbone network in the following models.

For concise presentation, we list key experimental results on the development test set from Table 1 to Table 3. Please note that, each utterance in the development set of the challenge is truncated or padded repeatedly to be 6-second long. In Table 1, we show the superior performance of the proposed system in comparisons with two baseline systems. B1 denotes the first baseline and B2 denotes the second baseline. We can see that, the challenge is quite difficult because the PESQ score of the first noisy channel is low. B2 model combines both time and frequency domain features with a TCN and obtained improvement against the challenge baseline B1. With our proposed model, the four kinds of measures were improved by a large margin with respect to B1 and B2, e.g. the PESQ score increases by 0.24 and 0.19.

Table 1: *Four kinds of objective measures of the proposed system in comparisons with two baselines.*

| model symbol | PESQ | STOI | E-STOI | SI-SNR |
|---|---|---|---|---|
| noisy | 1.515 | 0.823 | 0.691 | 4.475 |
| B1 | 1.638 | 0.840 | 0.706 | 6.789 |
| B2 | 1.679 | 0.856 | 0.732 | 7.414 |
| Proposed | **1.873** | **0.880** | **0.768** | **8.265** |

For our proposed two-stage feature fusion, the Conv2D layer is essential. In Table 2, the systems with and without the Conv2D fusion layer are compared. SC stands for single-channel. In the tables, intra-channel (intra-ch) feature means the feature is extracted within a channel, while we refer a spatial feature computed among different channels as an inter-channel (inter-ch) feature. The two systems in Table 2 both operate in the frequency domain. For model M1, the input single-channel LPS is just concatenated with the IPDs along the frequency domain and feed into a subsequent Conv1D layer, where the M2 model put the two kinds of features as different panels to be input into a Conv2D layer and then a subsequent Conv1D layer. With Conv2D fusion, M2 model obviously outperformed the M1 model. We also find that though time-domain features are not used in M2, M2 still gained obvious improvements compared to the B2 model. It demonstrates that Conv2d is quite useful for feature fusion. We then show in Table 3 how the proposed model was built with the two-stage fusion using three Conv2D layers.

Table 2: *Four kinds of objective measures of speech enhancement systems with and without a Conv2D fusion layer.*

| model symbol | intra-ch | inter-ch | fusion | PESQ | STOI | E-STOI | SI-SNR |
|---|---|---|---|---|---|---|---|
| M1 | SC LPS | IPD | - | 1.638 | 0.853 | 0.729 | 7.334 |
| M2 | SC LPS | IPD | Conv2D | 1.756 | 0.860 | 0.736 | 7.875 |

Table 3: *Four kinds of objective measures of four speech enhancement systems with different configurations.*

| model symbol | feature | | | | | fusion-3 | backbone | AM | PESQ | STOI | E-STOI | SI-SNR |
| | frequency-domain | | | time-domain | | | | | | | | |
| | intra-ch | inter-ch | fusion-1 | inter-ch | fusion-2 | | | | | | | |
|---|---|---|---|---|---|---|---|---|---|---|---|---|
| M3 | MC LPS | - | - | - | - | Conv2D | TCN | - | 1.773 | 0.869 | 0.749 | 8.050 |
| M4 | MC LPS | SDBF LPS | - | - | - | Conv2D | TCN | - | 1.769 | 0.870 | 0.753 | 8.122 |
| M5 | MC LPS | SDBF LPS | Conv2D | MCS+ICD | Conv2D | Conv2D | TCN | - | 1.779 | 0.871 | 0.754 | 8.093 |
| Proposed | MC LPS | SDBF LPS | Conv2D | MCS+ICD | Conv2D | Conv2D | TCN | TDNN | **1.873** | **0.880** | **0.768** | **8.265** |

In Table 3, four systems using multi-channel LPS features with different configurations are compared, which are all better than the M2 model fusing the single-channel LPS and IPDs. MC represents multi-channel, fusion-1 is the first fusion layer of the first fusion stage and fusion-2 means the second fusion layer of the first fusion stage, while fusion-3 denotes the third fusion in the second fusion stage. M3 denotes a system using one Conv2D layer fusing multi-channel LPS. M4 model fuse MC LPS and SDBF LPS together. M5 model uses two Conv2D fusion layers in the frequency and time domain in the first fusion stage separately and further combines them with the third fusion layer in the second stage. During the process, the objective measures kept becoming better as we fully utilized the temporal-frequency-spatial information from the array by applying the two-stage three fusion layers. Moreover, in our final proposed architecture, the acoustic model branch is used. The results show that with the acoustic model, all metrics get substantial improvements upon the M5 model, e.g. the PESQ score was improved by 0.094.

To further illustrate the progress made by out proposed model, we compare different processed spectrogram of Proposed and the two baselines in Fig. 3. As we can see, the first channel of the example utterance in Fig. 3a is quite reverberant and noisy. The first baseline can dereverberate and denoise to a limited extent in Fig. 3b, while the second baseline in Fig. 3c removed much reverberation and background noise but cancelled some formant of the original speech and produced a lot of distortion, especially in the medium and high frequency domain. Shown in Fig. 3d, our proposed model performed best, reducing the noise and reverberation as well as constraining the distortion, as highlighted in the green box.

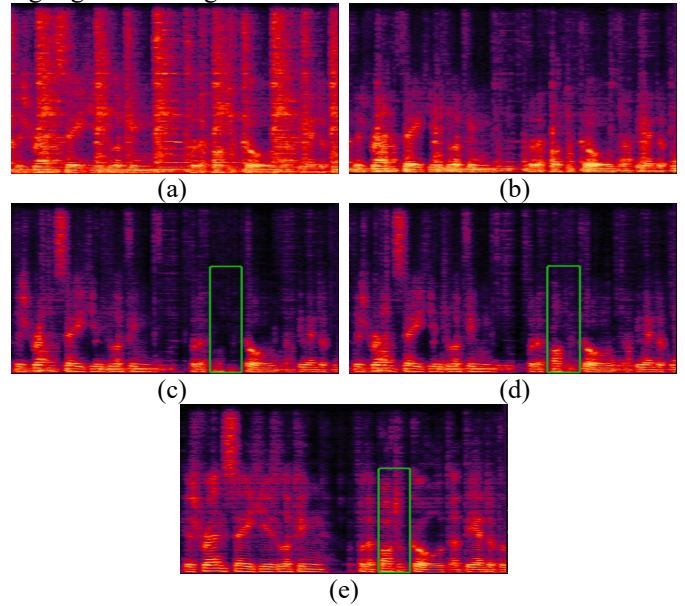

Fig. 3 Processed spectrograms of an example utterance. (a) The noisy first channel. (b) The official baseline output. (c) The second baseline model output. (d) The proposed model output. (e) Clean reference.

## IV. CONCLUSIONS

We have proposed a two-stage feature fusion approach for multi-channel enhancement and a novel loss function with a pre-trained acoustic model to constrain the speech distortion. In the feature fusion approach, we combine the multi-channel LPS features and SDBF features in the frequency domain using the first fusion Conv2D layer, incorporate the MCS with the ICD features in the time domain with a second fusion Conv2D layer, and further fuse the two above outputs with the third fusion Conv2D layer. In the acoustic model, a TDNN model is pre-trained with clean speech using the end-to-end LFMMI criterion. The results on the Task1 development test set of the ConferencingSpeech 2021 challenge demonstrate that our proposed model obtains the best performance on comparisons to the official baseline and a recently proposed method. We also observe obvious source speech restoration of the utterances processed by our model. In the near future, we will explore more on the distributed multi-array speech enhancement task assisted by ASR objectives.